\documentclass[smallextended]{svjour3}       % onecolumn (second format)
\smartqed  % flush right qed marks, e.g. at end of proof
\usepackage{graphicx}
\usepackage{amsmath}
\usepackage{amssymb}
\usepackage{hyperref}
\usepackage{color}
\usepackage{longtable}
\usepackage{blindtext}
\usepackage{rotating}
\usepackage{lscape}
\usepackage{caption}
\usepackage[numbers,sort&compress]{natbib}
\begin{document}

\title{Microscopic Investigation of Ground State Properties and Shape Evolution in Osmium Isotopes
}


\author{Usuf Rahaman        \and
	 M.Ikram        \and
	 Ishfaq~A.~Rather        \and 
	 Anisul Ain Usmani %etc.
}


\institute{Usuf Rahaman    \at
	Department of Physics, Madanapalle Institute of Technology $\&$ Science, Madanapalle-517325, India.
	\email{physics.usuf@gmail.com}  
	    \and
	M.Ikram     \at
	Department of Physics, Harsh Vidya Mandir (PG) College Raisi, Haridwar-247671, India.
%	\email{physics.usuf@gmail.com}  
	\and
	Ishfaq~A.~Rather      \at
%Centro de Astrof{\'i}sica e Gravita{\c c}{\~a}o-CENTRA, Instituto Superior T{\'e}cnico,   Universidade de Lisboa, 1049-001 Lisboa, Portugal.
Institut f\"{u}r Theoretische Physik, Goethe Universit\"{a}t, 60438 Frankfurt am Main, Germany.
	\and
	Anisul Ain Usmani    \at
	Department of Physics, Aligarh Muslim University, Aligarh-202002, India
}

\date{Received: date / Accepted: date}

\maketitle

\begin{abstract}
The present study focuses on investigating the shape evolution of neutron-rich even-even Osmium (Os) transitional nuclei within the range of neutron number N = 82 to N = 190. The investigation is conducted using density-dependent meson-nucleon and point-coupling models within the framework of the covariant density functional theory (CDFT). Additionally, the results obtained from the CDFT calculations are compared with those obtained using the relativistic mean-field model with a non-linear meson-nucleon interaction. The potential energy curve for Os isotopes (ranging from $^{158}$Os to $^{260}$Os) is analyzed in order to identify phase shape transitions, such as oblate-spherical-prolate. Furthermore, ground state bulk properties are calculated to gain insights into the structure of Os isotopes. The self-consistent calculations reveal a clear shape transition in the even-even Os isotopes, and overall, good agreement is observed among the different models employed as well as with the available experimental data.
\end{abstract}


%\ccode{PACS numbers:}

%\tableofcontents

\section{Introduction}
The investigation of transitional nuclei is of paramount importance for the comprehensive 
understanding of nuclear structural properties, particularly with regard to their 
prolate and oblate shaped configurations. These nuclei, characterized by their 
dynamic and static effects, have become a focal point of both theoretical and experimental studies. 
Over the past few decades, various experiments~\cite{alkh,whel,john,pt1,pt2} and theoretical models~\cite{stev,afana2016,afana2014,Sarriguren2008,rob2009,rod2010}, 
both relativistic and non-relativistic in nature, have been employed to 
study transitional nuclei with $ \mathrm{Z} = 72-80 $. A numerous number 
of studies have documented the evolution of triaxial shapes in these isotopes, 
further enriching our knowledge in this field~\cite{Nomura1,Nomura2,Nomura3,Ramos2014}.

Furthermore, extensive investigations have observed the presence of 
superdeformed nuclei in this specific region, as evidenced by their non-spherical 
shapes~\cite{Robert1991,Sun2009}. The existence of deformed shapes in nuclei 
disrupts the inherent spontaneous symmetry and introduces non-sphericity, 
primarily governed by the quadrupole moment. Notably, certain nuclei 
in this region exhibit multiple deformations at nearly identical excitation energies, 
indicating the occurrence of shape co-existence. This phenomenon allows for 
the exploration of nuclear oscillations between different shapes. 
These distinct characteristics make the transitional nuclei within 
this region highly promising candidates for in-depth structural studies. 
It is worth noting that the most intriguing area within the nuclear 
transition range is believed to lie around the mass number A$\sim$190, 
where the transformation between prolate and oblate shapes has been 
thoroughly investigated~\cite{john}. The region comprising transitional 
neutron-rich nuclei has also captured significant attention within 
the nuclear physics community due to its implications for neutron magicity~\cite{magic31,magic32,magic33}.

The isotopes of osmium play a significant role in the transitional region, 
where the nuclear shape undergoes a transition from spherical to prolate/oblate configurations, 
and subsequently returns to a spherical shape as the neutron number increases. 
Nuclei within this region provide a suitable framework for a critical examination of shape evolution. 
Specifically, we focus our investigation on the even-even osmium isotopes ($^{158-260}$Os), 
characterized by a high neutron count. Thus, our study expands our understanding 
from the beta stable region to the neutron-drip line region. 
It is important to note that pairing correlation is of utmost importance in these loosely bound nuclei, 
which exhibit intriguing phenomena such as neutron halo and skin~\cite{tanihata1996,horo2001,Horowitz2001}. 
Consequently, an explicit treatment of pairing correlations is essential when describing nuclei near the drip line.

Density Functional Theory (DFT) is a highly versatile and powerful approach employed 
to evaluate various physical observables of nuclei across the entire periodic table, 
utilizing both relativistic and non-relativistic
functionals~\cite{stev,afana2016,afana2014,Sarriguren2008,rob2009,rod2010,bartel1982,doba1984,chabanat1998,klupfel2009,kortel2010,kortel2012,erler2012}. 
This framework not only addresses the bulk properties but 
also enables the investigation of transitional behavior in nuclei. 
 Notably, few studies have reported calculations on nuclear 
 shape transitions employing relativistic functionals~\cite{afana2016,afana2014}. 
Additionally, self-consistent Hartree-Fock calculations based on Skyrme energy density functionals have been
conducted~\cite{Sarriguren2008,rob2009,rod2010,bartel1982,doba1984,chabanat1998,klupfel2009,kortel2010,kortel2012,erler2012}.
Furthermore, even-even osmium isotopes have been investigated within the framework of Hartree-Fock Bogoliubov using Skyrme interactions~\cite{Ashok2019}. 
Despite the considerable theoretical considerations in Covariant Density Functional Theory (CDFT), 
there remains a scarcity of research concerning shape evolution in transitional nuclei. 
Consequently, our current endeavor focuses on employing CDFT to explore the spherical 
to prolate/oblate shape transitions in osmium transitional nuclei. Our primary objective 
is to attain a more comprehensive and systematic understanding of these nuclear isotopes 
within both relativistic and non-relativistic functionals. 
The calculations entail Relativistic Mean-Field (RMF) theory~\cite{106,gambhir1990,7,walecka1974,BB77} 
incorporating nonlinear meson-nucleon NL3*~\cite{nl3s} interaction with BCS pairing, 
as well as covariant density functional theory (CDFT)~\cite{13,42} 
utilizing density-dependent meson-exchange (DD-ME)~\cite{me1,me2}
and density-dependent point-coupling interactions (DD-PC)~\cite{pc1,pcx} with RHB pairing. 

Within this paper, we employ these methodologies to provide valuable insights into the 
nuclear shape transitions exhibited in the even-even isotopic chain of osmium, 
encompassing the mass range from $^{158}$Os to $^{260}$Os. 
As previously indicated, this mass region offers a captivating opportunity for 
exploring neutron magic numbers and investigating small islands of phase shape transition, 
which could hold energetic favorability for reaction mechanisms. This investigation yields valuable 
insights into various nuclear properties, including ground state energy, 
quadrupole deformation, radii, nuclear skin, and potential energy curve. 
It is worth noting that while triaxial symmetry plays a crucial role in these nuclei, 
our study focuses exclusively on axially deformed cases.

The organization of this paper is as follows. 
Section 2 provides a comprehensive overview of the theoretical formalism employed, 
along with detailed explanations of the calculations employed in this study. 
The results of these calculations, along with their pertinent discussions, 
are presented in Section 3. Finally, Section 4 concludes the paper with a summary of the key findings.

\section{Theoretical Framework}

\subsection{Covariant density functional theory}
In this paper, we employ two classes of covariant density functional models, namely the density-dependent meson-exchange (DD-ME) model and the density-dependent point-coupling (DD-PC) model. These models differ primarily in their treatment of interaction range. The DD-ME model incorporates a finite interaction range, whereas the DD-PC model utilizes a zero-range interaction supplemented by an additional gradient term in the scalar-isoscalar channel. By employing these distinct models, we aim to comprehensively explore the impact of interaction range on the results and implications of our study. 
\subsubsection{The meson-exchange model}
Within the meson-exchange model, the nucleus is depicted as a system comprising Dirac nucleons that interact through the exchange of mesons with finite masses, resulting in interactions of finite range.~\cite{42,me1,me2,typel1999,52,111}.
These mesons, namely the isoscalar-scalar $\sigma$ meson, the isoscalar-vector $\omega$ meson, and the isovector-vector $\rho$ meson, constitute the essential set of meson fields required for a comprehensive quantitative description of nuclei. The meson-exchange model is characterized by the utilization of a standard Lagrangian density~\cite{gambhir1990} that incorporates vertices with medium-dependent interactions:
\begin{eqnarray}
	{\cal L}&=&\bar{\psi}\left[
	\gamma(i\partial-g_{\omega}\omega-g_{\rho
	}\vec{\rho}\,\vec{\tau}-eA)-m-g_{\sigma}\sigma\right]  \psi\nonumber\\
	&+&\frac{1}{2}(\partial\sigma)^{2}-\frac{1}{2}m_{\sigma}^{2}\sigma^{2}%
	-\frac{1}{4}\Omega_{\mu\nu}\Omega^{\mu\nu}+\frac{1}{2}m_{\omega}^{2}\omega
	^{2}\label{lagrangian}\nonumber\\
	&-&\frac{1}{4}{\vec{R}}_{\mu\nu}{\vec{R}}^{\mu\nu}+\frac{1}{2}m_{\rho}%
	^{2}\vec{\rho}^{\,2}-\frac{1}{4}F_{\mu\nu}F^{\mu\nu}.
\end{eqnarray}
In equation (\ref{lagrangian}), the symbol $\psi$ represents the Dirac spinors, while $m$ denotes the mass of the nucleon. The variable $e$ corresponds to the charge of the proton, which is zero for neutrons. The quantities $m_\sigma$, 
$m_\omega$, and $m_\rho$ represent the masses of the $\sigma$ meson, $\omega$ meson, 
and $\rho$ meson, respectively. The corresponding coupling constants for the interaction of these mesons with the nucleons are denoted as $g_\sigma$ , $g_\omega$, and $g_\rho$. These coupling constants, along with the unknown meson masses, serve as parameters in the Lagrangian equation. The symbols $\Omega^{\mu\nu}$, ${\vec{R}}^{\mu\nu}$ 
and $F^{\mu\nu}$ refer to the field tensors associated with the vector fields $\omega$, $\rho$ and the proton, respectively.
\begin{equation}
	\Omega^{\mu\nu}=\partial^{\mu}\omega^{\nu}-\partial^{\nu}\omega^{\mu},
\end{equation}
\begin{equation}
	{\vec{R}}^{\mu\nu}=\partial^{\mu}{\vec{\rho}}^{\nu}-\partial^{\nu}{\vec{\rho}}^{\mu},
\end{equation}
\begin{equation}
	F^{\mu\nu}=\partial^{\mu}A^{\nu}-\partial^{\nu}A^{\mu}.
\end{equation}
The initial formulation of this linear model was proposed by Walecka \cite{walecka1974}. However, this simplistic model lacks the capability to provide a quantitative depiction of nuclear systems ~\cite{BB77,Pannert1987} due to its linear interaction terms in the meson fields. Notably, it exhibits a considerably exaggerated incompressibility for infinite nuclear matter~\cite{BB77} and insufficiently small nuclear deformations~\cite{gambhir1990}. To address these limitations and achieve a more realistic portrayal of complex nuclear system properties, one can incorporate either a nonlinear self-coupling or density dependence in the coupling constants.

For the inclusion of nonlinear self-coupling, an additional term needs to be incorporated into the Lagrangian: 

\begin{equation}
	U(\sigma)~=~\frac{1}{2}m_{\sigma}^{2}\sigma^{2}+\frac{1}{3}g_{2}\sigma
	^{3}+\frac{1}{4}g_{3}\sigma^{4}.
\end{equation}
This model has demonstrated successful applications in various studies~\cite{reinhard1986,lalazissis1997,toddrutel2005}.

When it comes to density-dependent coupling constants, the dependence is defined as:
\begin{equation}
	g_{i}(\rho) = g_i(\rho_{\rm sat})f_i(x) \quad {\rm for} \quad i=\sigma, \omega, \rho
\end{equation}
Here, $i$ can represent any of the three mesons: $\sigma$, $\omega$ and $\rho$. Notably, there are no nonlinear terms present for the 
$\sigma$ meson, which implies that $g_2 = g_3 =0$. The density dependence or $\sigma$ and $\omega$ mesons are determined by the function:
\begin{equation}\label{fx}
	f_i(x)=a_i\frac{1+b_i(x+d_i)^2}{1+c_i(x+d_i)^2}.
\end{equation}
For the $\rho$ meson, the density dependence is given by:
\begin{equation}
	f_\rho(x)=\exp(-a_\rho(x-1)).
\end{equation}
In the above equations, \textit{x} represents the ratio between the baryonic density $\rho$ at a specific location and the baryonic density at saturation $\rho_{\rm sat}$ in symmetric nuclear matter. To ensure consistency, the parameters in  Eq.\ (\ref{fx}) are constrained as follows:
\begin{equation}
	f_i(1)=1,\,\,\,\, f_{\sigma}^{''}(1)=f_{\omega}^{''}(1),\,\, and \,\, f_{i}^{''}(0)=0.
\end{equation}
These constraints effectively reduce the number of independent parameters for the density dependence to three. In this study, the DD-ME1~\cite{me1} and DD-ME2~\cite{me2} density-dependent meson-exchange relativistic energy functionals are utilized, and their details can be found in Table \ref{table:1}.

%%%%%%%%%%%%%%%%%%%%%%%%%%%%%%%%%%%%%%%%%%%%%%%%%

\begin{table}[ht]
	\caption{The parameters of the DD-ME1~\cite{me1} and DD-ME2~\cite{me2} parameterizations of the Lagrangian} % title of Table
	\centering % used for centering table
	\begin{tabular}{c c c} % centered columns (4 columns)
		\hline\hline %inserts double horizontal lines
		Parameter &  DD-ME1~\cite{me1} &  DD-ME2~\cite{me2} \\ [0.5ex] % inserts table
		%heading
		\hline % inserts single horizontal line
		$m$		            &	939	        &	939 \\
		$m_{\sigma}$		&	549.5255	&	550.1238 \\
		$m_{\omega}$		&	783        	&	783	\\
		$m_{\rho}$	    	&	763     	&	763	\\
		$g_{\sigma}$		&	10.4434	&	10.5396	\\
		$g_{\omega}$		&	12.8939	&	13.0189	\\
		$g_{\rho}$		    &   3.8053	&	3.6836	\\
		$a_{\sigma}$		&	1.3854	&	1.3881	\\
		$b_{\sigma}$		&	0.9781	&	1.0943	\\
		$c_{\sigma}$		&	1.5342	&	1.7057	\\
		$d_{\sigma}$		&	0.4661	&	0.4421	\\
		$a_{\omega}$		&	1.3879	&	1.3892	\\
		$b_{\omega}$		&	0.8525	&	0.924	\\
		$c_{\omega}$		&	1.3566	&	1.462	\\
		$d_{\omega}$		&	0.4957	&	0.4775	\\
		$a_{\rho}$		    &	0.5008	&	0.5647	\\ [1ex] % [1ex] adds vertical space
		\hline
		\hline
	\end{tabular}
	\label{table:1} % is used to refer this table in the text
\end{table}
%%%%%%%%%%%%%%%%%%%%%%%%%%%%%%%%%%%%%%%%%%%%%%%%%

\subsubsection{Point-coupling model}
The Lagrangian associated with the density-dependent point coupling model~\cite{pc1,nikolaus1992} can be expressed as\\ 
\begin{eqnarray}
	{\cal L}&=&\bar{\psi}\left(i\gamma \cdot \partial-m\right)  \psi\nonumber\\
	&-&\frac{1}{2}\alpha_S(\hat\rho)\left(\bar{\psi}\psi\right)\left(\bar{\psi}\psi\right)
	-\frac{1}{2}\alpha_V(\hat\rho)\left(\bar{\psi}
	\gamma^{\mu}\psi\right)\left(\bar{\psi}\gamma_{\mu}\psi\right)\nonumber\\
	&-&\frac{1}{2}\alpha_{TV}(\hat\rho)\left(\bar{\psi}\vec\tau\gamma^{\mu}\psi\right)
	\left(\bar{\psi}\vec\tau\gamma_{\mu}\psi\right)\nonumber\\
	&-&\frac{1}{2}\delta_S\left(\partial_v\bar{\psi}\psi\right)
	\left(\partial^v\bar{\psi}\psi\right) - e\bar\psi\gamma \cdot A
	\frac{(1 - \tau_3)}{2}\psi.\label{Lag-pc}
\end{eqnarray}
The Lagrangian encompasses various components crucial for the density-dependent point coupling model. It incorporates the free-nucleon Lagrangian, point coupling interaction terms, and the coupling between protons and the electromagnetic field. The derivative terms present in Eq.\ (\ref{Lag-pc}) serve to account for the dominant effects originating from finite-range interactions, which hold significance in the context of nuclei. Similar to DD-ME models,  this particular formulation incorporates interactions involving isoscalar-scalar, isoscalar-vector, and isovector-vector couplings. 
In this investigation, the density-dependent point-coupling interactions, namely DD-PC1~\cite{pc1} and DD-PCX~\cite{pcx}, were adopted. These interactions are presented in Table \ref{table:2}.

%%%%%%%%%%%%%%%%%%%%%%%%%%%%%%%%%%%%%%%%%%%%%%%%%%%%%%

\begin{table}[ht]
	\caption{The parameters of the DD-PC1~\cite{pc1}, and DD-PCX~\cite{pcx} parameterization in the Lagrangian.}
	% title of Table
	\centering
	% used for centering table
	\begin{tabular}{c c c }
		% centered columns (3 columns)
		\hline
		\hline
		%inserts double horizontal lines
		
		Parameter &	DD-PC1~\cite{pc1} & DD-PCX~\cite{pcx} \\ [0.5ex]
		\hline
		$\it {m} $ (MeV)	&	939	&	939	\\
		$a_{s}$ (fm$^{2}$)      	&	-10.04616	&	-10.979243836	\\
		$b_{s}$ (fm$^{2}$)      	&	-9.15042	&	-9.03825091	\\
		$c_{s}$ (fm$^{2}$)      	&	-6.42729	&	-5.31300882	\\
		$d_{s}$                 	&	1.37235	&	1.37908707	\\
		$a_{v}$ (fm$^{2}$)      	&	5.91946	&	6.430144908	\\
		$b_{v}$ (fm$^{2}$)      	&	8.8637	&	8.870626019	\\
		$d_{v}$                 	&	0.65835	&	0.655310525	\\
		$b_{tv}$ (fm$^{2}$)     	&	1.83595	&	2.963206854	\\
		$d_{tv}$                	&	0.64025	&	1.309801417	\\
		$\delta_{s}$ (fm$^{4}$) 	&	-0.8149	&	-0.878850922	\\
		
		\hline
		\hline
	\end{tabular}
	\label{table:2} % is used to refer this table in the text
\end{table}
%%%%%%%%%%%%%%%%%%%%%%%%%%%%%%%%%%%%%%%%%%%%%%%%%%%%%%%

%%%%%%%%%%%%%%%%%%%%%%%%%%%%%%%%%%%%%%%%%%%%%%%%%
%%%%%%%%%%%%%%%%%%%%%%%%%%%%%%%%%%%%%%%%%%%%%%%%%%%%
\subsection{Axially deformed relativistic mean-field model}
The relativistic mean field (RMF) theory has emerged as a highly successful approach in addressing the nuclear many-body problem and elucidating various nuclear phenomena across the entire range of the periodic table, from the proton drip line to the neutron drip line region~\cite{S92,gambhir1990,R96,7,BB77,yadav,skpatra,Lalazissis1998}. One notable advantage of the RMF theory, compared to its non-relativistic counterpart, is its inherent ability to naturally reproduce the spin-orbit splitting. This feature plays a pivotal role in explaining the formation of closed shell structures in atomic nuclei~\cite{sharma,warrier}.

The foundation of the RMF theory lies in a fundamental Lagrangian density that encompasses nucleons interacting with  $\sigma-$, $\omega-$ and $\rho-$meson fields. Additionally, the inclusion of the photon field $A_{\mu}$ is crucial for adequately accounting for the Coulomb interaction among protons. The Lagrangian density of the relativistic mean field theory can be expressed as~\cite{gambhir1990,7,BB77,R96,S92},
\begin{eqnarray}
	{\cal L}&=&\bar{\psi_{i}}\{i\gamma^{\mu}
	\partial_{\mu}-M\}\psi_{i}
	+{\frac12}\partial^{\mu}\sigma\partial_{\mu}\sigma
	-{\frac12}m_{\sigma}^{2}\sigma^{2}
	-{\frac13}g_{2}\sigma^{3} \nonumber \\
	&-&{\frac14}g_{3}\sigma^{4}-g_{s}\bar{\psi_{i}}\psi_{i}\sigma 
	-{\frac14}\Omega^{\mu\nu}\Omega_{\mu\nu}+{\frac12}m_{w}^{2}V^{\mu}V_{\mu}\nonumber \\
	&-&g_{w}\bar\psi_{i}\gamma^{\mu}\psi_{i}
	V_{\mu}-{\frac14}\vec{B}^{\mu\nu}\vec{B}_{\mu\nu} 
	+{\frac12}m_{\rho}^{2}{\vec{R}^{\mu}}{\vec{R}_{\mu}}-{\frac14}F^{\mu\nu}F_{\mu\nu} \nonumber\\
	&-&g_{\rho}\bar\psi_{i}\gamma^{\mu}\vec{\tau}\psi_{i}\vec{R^{\mu}}-e\bar\psi_{i}
	\gamma^{\mu}\frac{\left(1-\tau_{3i}\right)}{2}\psi_{i}A_{\mu} .
\end{eqnarray}
Here, the masses of the nucleons, ${\sigma}$-, ${\omega}$- and ${\rho}$-mesons are represented by $M$, $m_{\sigma}$, $m_{\omega}$ and $m_{\rho}$, respectively. The fields associated with the ${\sigma}$-, ${\omega}$- and ${\rho}$-mesons are denoted by ${\sigma}$, $V_{\mu}$ and $R_{\mu}$. The coupling constants for the ${\sigma}$-, ${\omega}$- and ${\rho}$-mesons, as well as the electromagnetic interaction, are given by $g_s$, $g_{\omega}$, $g_{\rho}$ and $e^2/4{\pi}$=1/137, respectively. Additionally, the self-interaction coupling constants for ${\sigma}$-mesons are denoted by $g_2$ and $g_3$.

By employing the classical variational principle, the field equations governing the behavior of nucleons and mesons can be derived. Specifically, the Dirac equation for nucleons, which encapsulates their dynamics, can be written as
\begin{equation}
	\{-i\alpha\bigtriangledown + V(r_{\perp},z)+\beta M^\dagger\}\psi_i=\epsilon_i\psi_i.
\end{equation}
Where
\begin{equation}
	M^\dagger=M+S(r_{\perp},z)=M+g_\sigma\sigma^0(r_{\perp},z),
\end{equation}
and 
\begin{equation}
	V(r_{\perp},z)=g_{\omega}V^{0}(r_{\perp},z)+g_{\rho}\tau_{3}R^{0}(r_{\perp},z)+
	e\frac{(1-\tau_3)}{2}A^0(r_{\perp},z). 
\end{equation}
The Klein-Gordon equations for mesons are given by 
\begin{eqnarray}
	\{-\bigtriangleup+m^2_\sigma\}\sigma^0(r_{\perp},z)&=&-g_\sigma\rho_s(r_{\perp},z)\nonumber\\
	&-& g_2\sigma^2(r_{\perp},z)-g_3\sigma^3(r_{\perp},z) ,\\
	\{-\bigtriangleup+m^2_\omega\}V^0(r_{\perp},z)&=&g_{\omega}\rho_v(r_{\perp},z) ,\\
	\{-\bigtriangleup+m^2_\rho\}R^0(r_{\perp},z)&=&g_{\rho}\rho_3(r_{\perp},z) , \\
	-\bigtriangleup A^0(r_{\perp},z)&=&e\rho_c(r_{\perp},z). 
\end{eqnarray}
The scalar and vector densities for the $\sigma$- and $\omega$-fields are denoted as $\rho_s$ and $\rho_v$, respectively. These densities represent the distribution and magnitude of the $\sigma$-meson and $\omega$-meson fields within the nuclear environment. Mathematically, these densities are expressed as 

\begin{eqnarray} 
	\rho_s(r) &=& \sum_ {i=n,p}\bar\psi_i(r)\psi_i(r)\;,                         \nonumber \\
	\rho_v(r) &=& \sum_{i=n,p}\psi^\dag_i(r)\psi_i(r) \;.			      
\end{eqnarray}
The vector density $\rho_3(r)$ for $\rho$-field and charge density 
$\rho_c(r)$ are expressed by 
\begin{eqnarray} 
	\rho_3(r) &=& \sum_{i=n,p} \psi^\dag_i(r)\gamma^0\tau_{3i}\psi_i(r)\; ,		      \nonumber \\
	\rho_c(r) &=& \sum_{i=n,p} \psi^\dag_i(r)\gamma^0\frac{(1-\tau_{3i})}{2}\psi_i(r)\;.    
\end{eqnarray}
To describe the fundamental characteristics of nuclei in their ground state, a static solution is obtained by solving the equations of motion. These equations, which are nonlinear and coupled, capture the interactions and dynamics of nucleons and mesons within the nuclear system. The solution is achieved through a self-consistent approach, where the equations are solved iteratively to ensure internal consistency.

In this particular study, an axially deformed harmonic oscillator basis is employed for both fermions and bosons. The basis size is chosen as $N_F = N_B = 20$, indicating the number of states included in the basis for fermions and bosons, respectively. 

One of the important quantities of interest in this study is the calculation of radii. These radii are determined from the corresponding densities obtained from the solution. The densities provide information about the spatial distribution and size of the nuclear system, and the radii serve as quantitative measures of these distributions. The rms radii of proton ($r_p$), neutron ($r_n$) and nuclear matter ($r_m$) are given by,
\begin{eqnarray}
	\langle r_p^2\rangle &=& \frac{1}{Z}\int r_p^{2}d^{3}r\rho_p\;,        \nonumber \\
	\langle r_n^2\rangle &=& \frac{1}{N}\int r_n^{2}d^{3}r\rho_n\;,        \nonumber \\
	\langle r_m^2\rangle &=& \frac{1}{A}\int r_m^{2}d^{3}r\rho\;,          
\end{eqnarray} 
The rms charge radius of a nucleus can be determined by utilizing the proton rms radius and accounting for the finite size of the proton. This is achieved by employing the relation $r_c = \sqrt{r_p^2 + 0.64}$.

To extract the quadrupole deformation parameter ($\beta_2$) of the nucleus, the calculated quadrupole moments of neutrons and protons are used as
\begin{equation}
	Q = Q_n + Q_p = \sqrt{\frac{16\pi}5} \left(\frac3{4\pi} AR^2\beta_2\right),
\end{equation}
where R = 1.2$A^{1/3}$.\\
The total energy of the nuclear system is composed of various contributions.
\begin{equation}
	E_{\text{total}} = E_{\text{part}}+E_{\sigma}+E_{\omega}+E_{\rho}+E_{c}+E_{\text{pair}}+E_{c.m.},
\end{equation}
Firstly, the sum of single particle energies of the nucleons is denoted as $E_{\text{part}}$. Additionally, there are contributions from the meson fields, namely $E_{\sigma}$, $E_{\omega}$, and $E_{\rho}$. The Coulomb field contributes to the energy as $E_{c}$, and the pairing energy is represented by $E_{\text{pair}}$. Furthermore, the center-of-mass energy is denoted as $E_{\text{cm}}$.

The non-linear NL3* parameter set~\cite{nl3s}, which is a modified version of the successful NL3 parameter set~\cite{lalazissis1997}, is utilized throughout the calculations. This parameter set provides the necessary parameters for the model being employed.
For a detailed understanding of the formalism and numerical techniques used, references such as Refs.~\cite{gambhir1990,patra} and other relevant literature should be consulted.

In this study, pairing correlations are treated using the BCS (Bardeen-Cooper-Schrieffer) approach~\cite{Dobaczewski1984}. Although the BCS approach may not be appropriate for light neutron-rich nuclei, the nuclei considered in this study are not in that category, ensuring the reliability of the results obtained within the framework of the relativistic mean field (RMF) theory.

%%%%%%%%%%%%%%%%%%%%%%%%%%%%%%%%%%%%%%%%%%%%%%%%%%%%%%%

\begin{table}[ht]
	\caption{The parameters of the NL3*~\cite{nl3s} parameterization of the Lagrangian} % title of Table
	\centering % used for centering table
	\begin{tabular}{c c} % centered columns (4 columns)
		\hline\hline %inserts double horizontal lines
		Parameter &  NL3*~\cite{nl3s} \\ [0.5ex]
		\hline
		$m$   & 939  \\
		$m_{\sigma}$  & 502.5742 \\
		$m_{\omega}$  & 782.6  \\
		$m_{\rho}$    & 763 \\
		$g_{\sigma}$  & 10.0944 \\
		$g_{\omega}$  & 12.8065 \\
		$g_{\rho}$    &  4.5748 \\
		$g_{2}$       & -10.8093 \\
		$g_{3}$       & -30.1486 \\			
		\hline \hline
	\end{tabular}
	\label{table:3} % is used to refer this table in the text
\end{table}

%%%%%%%%%%%%%%%%%%%%%%%%%%%%%%%%%%%%%%%%%%%%%%%%%%%%%%%%
%RESULTS AND DISCUSSIONS
%%%%%%%%%%%%%%%%%%%%%%%%%%%%%%%%%%%%%%%%%%%%%%%%%%%%%%%%
\section{Results and Discussion}

\subsection{Binding energy}

\begin{figure}[hbt!]
	\begin{center}
		\includegraphics[scale=0.45]{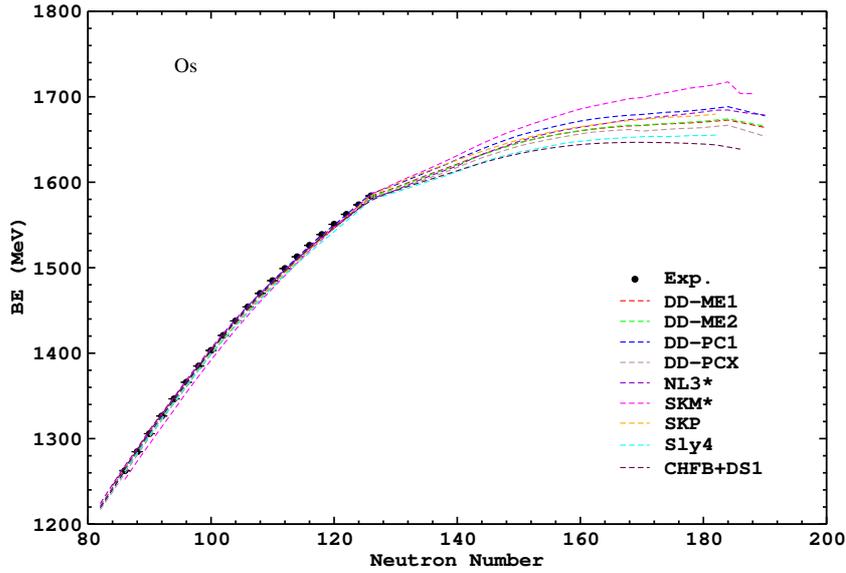}
		\caption{Total binding energy for even-even Os nuclei, 
			obtained within different relativistic interactions. 
		}
		\label{Osbe}
	\end{center}
\end{figure}

The binding energy is a crucial parameter for studying the stability of atomic nuclei. The binding energies of ground states for even-even osmium isotopes $^{160-264}$Os 
is shown in Table~\ref{tab:Ostab1} and plotted in Fig.~\ref{Osbe}. 
These energies are calculated within the CDFT with effective density-dependent interactions DD-ME1~\cite{me1}, 
DD-ME2~\cite{me2}, DD-PC1~\cite{pc1}, and DD-PCX~\cite{pcx} and also with RMF+BCS model 
using nonlinear NL3*~\cite{nl3s}.

\begin{figure}[hbt!]
	\begin{center}
		\includegraphics[scale=0.45]{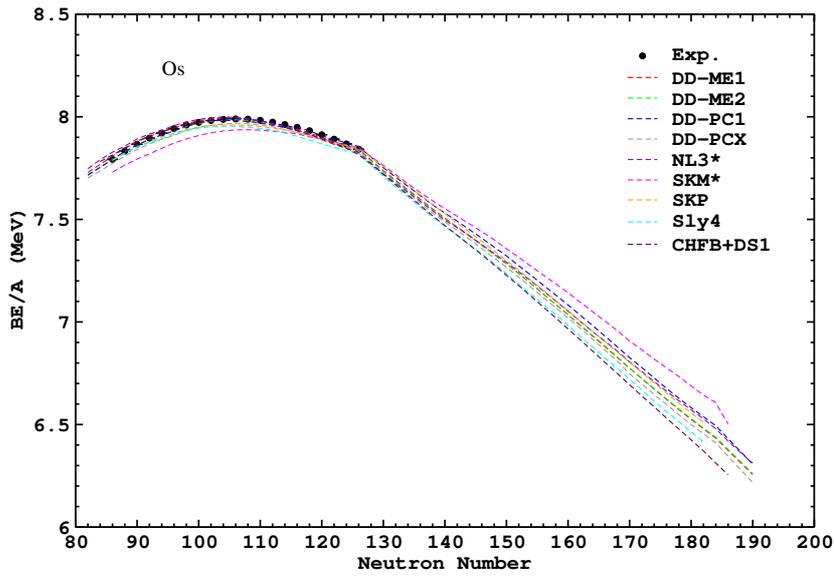}
		\caption{Binding energy per nucleon for even-even Os nuclei, obtained within different relativistic interactions. 
		}
		\label{Osbepn}
	\end{center}
\end{figure}

Additionally, Fig.~\ref{Osbepn} depicts the calculated binding energies for these isotopes. Since experimental data for neutron-rich drip-line nuclei is limited, it is necessary to compare the results obtained from different theoretical approaches~\cite{19,39} in order to assess the model's predictability and reliability. Comparisons were made with other theoretical frameworks, the HFB + THO approach in conjunction with various nuclear interactions, including Skyrme SLy4, SkP, and SkM*~\cite{19}, alongside CHFB+5DCH calculations predicated on the Gogny D1S interaction~\cite{39}, which exhibited excellent agreement with both RMF and CDFT outcomes, extending towards the neutron drip-line region.
The calculated binding energies exhibit a consistent agreement across the entire isotopic range. Particularly, the results obtained using the SkP functional demonstrate a strong alignment with our findings, extending even to the neutron dripline region. Conversely, the SkM* investigations indicate that isotopes in the neutron-rich region are slightly more tightly bound than our predictions, whereas the SLy4 and CHFB data suggest the opposite trend. The results obtained within the study range displayed good consistency among various relativistic and non-relativistic functionals, as well as with available experimental data~\cite{43,wang2012,22,Audi2003}.

In general, the binding energy per nucleon (BE/A) increases as the number of neutrons (or the mass number A) increases, reaching a maximum value at A = 182 for nearly all interactions. This trend is consistent with experimental observations. Consequently, it is concluded that the isotope $^{182}$Os is the most stable among the entire isotopic chain.

%%%%%%%%%%%%%%%%%%%%%%%%%%%%%%%%%%%%%%%%%%%%%%%%%%%%%%%55
\begin{landscape}
	
	\begin{longtable}{ccccccccccc}
		\caption{Comparison of the total binding energies (in MeV) of the ground states for $^{158-266}$Os isotopes calculated using different effective interactions (DD-ME1~\cite{me1}, DD-ME2~\cite{me2}, DD-PC1~\cite{pc1}, DD-PCX~\cite{pcx}, and NL3*~\cite{nl3s}) and with available experimental data~\cite{22, Audi2003, wang2012, 43}. The results are also compared with calculations using HFB models.~\cite{19,39}}\label{tab:Ostab1}\\
		%    	\toprule
		\small
		\textrm{Nuclei}&
		\textrm{DD-ME1}&
		\textrm{DD-ME2}&
		\textrm{DD-PC1}&
		\textrm{DD-PCX}&
		\textrm{NL3*}&
		\textrm{EXP}&
		\textrm{SkP}&
		\textrm{SkM*}&
		\textrm{SLy4}&
		\textrm{CHFB}\\
		\hline   
		$^{156}$Os 	&	1190.90	&	1188.57	&	1188.82	&	1186.16	&	1195.96	&	 --   	&	 -- 	&	 -- 	&	 -- 	&	--	\\
		$^{158 }$Os 	&	1221.14	&	1218.82	&	1219.25	&	1216.94	&	1224.09	&	 --   	&	 -- 	&	 -- 	&	 -- 	&	--	\\
		$^{160 }$Os 	&	1242.88	&	1240.44	&	1240.66	&	1237.69	&	1246.55	&	--	&	--	&	--	&	--	&	1245.14	\\
		$^{162 }$Os 	&	1265.02	&	1262.52	&	1262.49	&	1259.15	&	1268.13	&	1262.47	&	1252.25	&	1261.62	&	1261.88	&	1265.90	\\
		$^{164 }$Os 	&	1286.78	&	1284.22	&	1284.14	&	1280.62	&	1288.56	&	1284.66	&	1273.42	&	1282.98	&	1283.02	&	1286.93	\\
		$^{166 }$Os 	&	1307.78	&	1305.18	&	1305.20	&	1301.58	&	1310.12	&	1305.81	&	1293.73	&	1303.71	&	1303.37	&	1308.28	\\
		$^{168 }$Os 	&	1328.00	&	1325.39	&	1325.49	&	1321.88	&	1329.82	&	1326.52	&	1313.80	&	1323.77	&	1322.99	&	1328.20	\\
		$^{170 }$Os 	&	1347.54	&	1344.97	&	1345.12	&	1341.43	&	1348.75	&	1346.59	&	1333.75	&	1343.28	&	1341.93	&	1347.64	\\
		$^{172 }$Os 	&	1366.34	&	1363.84	&	1364.13	&	1360.42	&	1368.62	&	1366.05	&	1353.94	&	1362.30	&	1361.27	&	1366.82	\\
		$^{174 }$Os 	&	1386.15	&	1383.84	&	1384.43	&	1380.52	&	1387.63	&	1384.95	&	1373.24	&	1380.82	&	1380.18	&	1385.35	\\
		$^{176 }$Os 	&	1404.50	&	1402.21	&	1402.98	&	1398.95	&	1405.72	&	1403.19	&	1391.97	&	1399.12	&	1398.38	&	1403.27	\\
		$^{178 }$Os 	&	1422.06	&	1419.87	&	1420.77	&	1416.50	&	1423.13	&	1420.78	&	1410.08	&	1416.62	&	1415.36	&	1420.48	\\
		$^{180 }$Os 	&	1439.12	&	1437.10	&	1438.14	&	1433.66	&	1439.83	&	1437.74	&	1427.62	&	1432.87	&	1431.77	&	1436.86	\\
		$^{182 }$Os 	&	1455.14	&	1453.24	&	1454.54	&	1450.31	&	1455.48	&	1454.13	&	1444.31	&	1448.76	&	1447.59	&	1452.62	\\
		$^{184 }$Os 	&	1470.02	&	1468.38	&	1469.83	&	1465.75	&	1470.21	&	1469.92	&	1460.47	&	1464.21	&	1462.69	&	1467.84	\\
		$^{186 }$Os 	&	1484.24	&	1482.67	&	1484.35	&	1480.52	&	1484.21	&	1484.81	&	1475.92	&	1479.21	&	1477.43	&	1482.32	\\
		$^{188 }$Os 	&	1497.97	&	1496.49	&	1498.37	&	1494.75	&	1497.51	&	1499.09	&	1490.77	&	1493.75	&	1491.72	&	1496.23	\\
		$^{190 }$Os 	&	1511.24	&	1509.79	&	1512.04	&	1508.03	&	1510.40	&	1512.80	&	1505.47	&	1507.80	&	1505.03	&	1510.03	\\
		$^{192 }$Os 	&	1524.36	&	1523.05	&	1525.58	&	1521.39	&	1523.30	&	1526.12	&	1520.10	&	1521.30	&	1517.78	&	1523.00	\\
		$^{194 }$Os 	&	1537.08	&	1535.96	&	1538.79	&	1534.63	&	1535.59	&	1538.81	&	1532.93	&	1534.25	&	1530.09	&	1535.38	\\
		$^{196 }$Os 	&	1548.13	&	1546.99	&	1549.80	&	1545.42	&	1546.58	&	1550.80	&	1546.97	&	1546.91	&	1542.18	&	1547.07	\\
		$^{198 }$Os 	&	1559.81	&	1558.63	&	1562.16	&	1557.54	&	1557.50	&	1562.42	&	1559.72	&	1559.34	&	1554.55	&	1558.74	\\
		$^{200 }$Os 	&	1571.93	&	1570.99	&	1574.68	&	1570.10	&	1569.11	&	1573.60	&	1572.78	&	1571.95	&	1567.04	&	1569.47	\\
		$^{202 }$Os 	&	1584.15	&	1583.26	&	1586.67	&	1582.11	&	1580.48	&	1584.08	&	1585.78	&	1584.52	&	1579.54	&	1578.59	\\
		$^{204 }$Os 	&	1588.93	&	1587.96	&	1591.75	&	1586.29	&	1585.59	&	 --   	&	1592.33	&	1591.39	&	1583.78	&	1585.21	\\
		$^{206 }$Os 	&	1593.82	&	1592.82	&	1596.97	&	1590.65	&	1590.68	&	 --   	&	1599.08	&	1598.10	&	1588.35	&	1590.07	\\
		$^{208 }$Os 	&	1599.51	&	1598.52	&	1602.80	&	1595.96	&	1596.84	&	 --   	&	1605.54	&	1604.32	&	1593.12	&	1595.06	\\
		$^{210 }$Os 	&	1605.24	&	1604.25	&	1608.80	&	1601.36	&	1602.87	&	 --   	&	1611.54	&	1610.04	&	1597.95	&	1599.77	\\
		$^{212 }$Os 	&	1610.77	&	1609.79	&	1614.62	&	1606.59	&	1608.61	&	 --   	&	1617.86	&	1615.30	&	1602.71	&	1604.35	\\
		$^{214 }$Os 	&	1615.99	&	1615.04	&	1620.27	&	1611.64	&	1613.94	&	 --   	&	1624.44	&	1620.69	&	1607.33	&	1608.89	\\
		$^{216 }$Os 	&	1621.26	&	1620.51	&	1626.77	&	1617.35	&	1619.07	&	 --   	&	1631.21	&	1625.98	&	1612.50	&	1613.49	\\
		$^{218 }$Os 	&	1627.01	&	1626.37	&	1633.04	&	1623.12	&	1626.20	&	 --   	&	1638.37	&	1631.13	&	1617.79	&	1618.23	\\
		$^{220 }$Os 	&	1632.42	&	1631.93	&	1638.97	&	1628.56	&	1632.46	&	 --   	&	1645.16	&	1636.12	&	1623.10	&	1622.56	\\
		$^{222 }$Os 	&	1637.33	&	1636.88	&	1644.47	&	1633.31	&	1637.72	&	 --   	&	1651.58	&	1640.93	&	1627.78	&	1626.58	\\
		$^{224 }$Os 	&	1641.91	&	1641.56	&	1649.75	&	1637.93	&	1642.84	&	 --   	&	1657.50	&	1645.53	&	1631.78	&	1630.19	\\
		$^{226 }$Os 	&	1646.11	&	1645.91	&	1654.68	&	1642.38	&	1647.76	&	 --   	&	1662.68	&	1649.70	&	1635.18	&	1633.41	\\
		$^{228 }$Os 	&	1649.73	&	1649.59	&	1658.58	&	1645.67	&	1651.93	&	 --   	&	1667.79	&	1653.45	&	1637.98	&	1636.26	\\
		$^{230 }$Os 	&	1652.87	&	1652.81	&	1662.17	&	1648.72	&	1655.84	&	 --   	&	1672.43	&	1656.77	&	1640.77	&	1638.72	\\
		$^{232 }$Os 	&	1655.57	&	1655.63	&	1665.50	&	1651.54	&	1659.12	&	 --   	&	1676.94	&	1659.83	&	1643.70	&	1640.82	\\
		$^{234 }$Os 	&	1657.97	&	1658.14	&	1668.65	&	1654.15	&	1661.80	&	 --   	&	1681.53	&	1662.57	&	1646.51	&	1642.62	\\
		$^{236 }$Os 	&	1660.15	&	1660.46	&	1671.60	&	1656.67	&	1664.31	&	 --   	&	1685.71	&	1664.93	&	1647.93	&	1644.08	\\
		$^{238 }$Os 	&	1661.99	&	1662.41	&	1673.95	&	1658.61	&	1666.48	&	 --   	&	1688.90	&	1667.03	&	1649.45	&	1645.18	\\
		$^{240 }$Os 	&	1663.52	&	1664.06	&	1675.72	&	1659.90	&	1668.69	&	 --   	&	1691.94	&	1668.93	&	1650.62	&	1645.92	\\
		$^{242 }$Os 	&	1664.84	&	1665.50	&	1677.23	&	1660.92	&	1671.16	&	 --   	&	1694.86	&	1670.64	&	1651.56	&	1646.40	\\
		$^{244 }$Os 	&	1665.83	&	1666.62	&	1678.42	&	1661.63	&	1673.39	&	 --   	&	1697.80	&	1672.13	&	1652.42	&	1646.64	\\
		$^{246 }$Os 	&	1666.37	&	1666.70	&	1679.32	&	1659.78	&	1673.96	&	 --   	&	1699.14	&	1673.52	&	1653.30	&	1646.64	\\
		$^{248 }$Os 	&	1667.41	&	1667.86	&	1680.58	&	1660.86	&	1676.05	&	 --   	&	1702.17	&	1674.74	&	1653.70	&	1646.45	\\
		$^{250 }$Os 	&	1668.14	&	1668.70	&	1681.87	&	1661.85	&	1677.29	&	 --   	&	1704.83	&	1675.79	&	1653.35	&	1646.40	\\
		$^{252 }$Os 	&	1668.73	&	1669.40	&	1682.47	&	1662.74	&	1678.96	&	 --   	&	1707.58	&	1676.59	&	1653.85	&	1645.91	\\
		$^{254 }$Os 	&	1669.50	&	1670.41	&	1683.59	&	1663.19	&	1680.47	&	 --   	&	1710.32	&	1677.30	&	1654.60	&	1645.38	\\
		$^{256 }$Os 	&	1670.38	&	1671.36	&	1685.00	&	1664.01	&	1682.30	&	 --   	&	1712.08	&	1678.31	&	1654.66	&	1644.68	\\
		$^{258 }$Os 	&	1671.52	&	1672.69	&	1686.70	&	1665.42	&	1684.30	&	 --   	&	1714.25	&	1679.77	&	1655.17	&	1643.74	\\
		$^{260 }$Os 	&	1672.75	&	1674.18	&	1688.43	&	1666.65	&	1684.52	&	 --   	&	1717.44	&	 -- 	&	 -- 	&	1641.23	\\
		$^{262 }$Os 	&	1669.92	&	1671.30	&	1684.88	&	1662.41	&	1682.07	&	 --   	&	1703.71	&	 -- 	&	 -- 	&	1638.69	\\
		$^{264 }$Os 	&	1667.03	&	1668.37	&	1681.31	&	1658.08	&	1680.20	&	 --   	&	1703.65	&	 -- 	&	 -- 	&	--	\\
		$^{266 }$Os 	&	1664.10	&	1665.39	&	1677.72	&	1653.66	&	1678.60	&	 --   	&	 --   	&	 -- 	&	 -- 	&	--\\
		\hline
	\end{longtable}
\end{landscape}
%\end{table*}

%%%%%%%%%%%%%%%%%%%%%%%%%%%%%%%%%%%%%%%%%5

%%%%%%%%%%%%%%%%%%%%%%%%%%%%%%%%%%%

\subsection{Quadrupole deformation}
\begin{figure}%[hbt!]
	\begin{center}
		\includegraphics[scale=0.45]{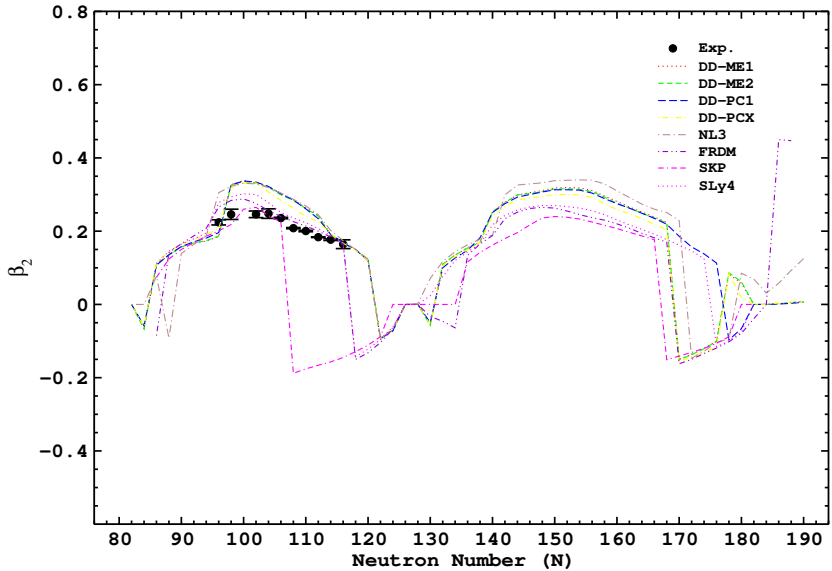}		
		\caption{The quadrupole deformation ($\beta_2$) of even-even Osmium nuclei from N=82 to neutron drip line, as obtained using different interactions.}
		\label{Osbeta}
	\end{center}
\end{figure}

Nuclear shape and deformation are significant physical parameters that 
play a crucial role in determining properties such as nuclear size, isotopic shift, and quadrupole moments. 
Table~\ref{tab:Osbeta} provides the quadrupole deformation parameters $\beta_2$) 
obtained from various relativistic and non-relativistic interactions. 
To provide a more clear understanding of the shape transition, 
Fig.~\ref{Osbeta} presents the variation deformation parameter with neutron number.

In order to assess the shape transition more comprehensively, 
we have compared the results obtained from the Relativistic Mean Field (RMF) 
approach with those obtained from non-relativistic functionals such as Skyrme 
SkP, SLy4, and SkM*~\cite{19}, as well as CHFB+5DCH~\cite{39} 
calculations based on the Gogny D1S interaction. 
These results were also compared with available experimental data~\cite{33,34}.
Throughout the considered isotopic chain, spanning from the neutron-deficient 
to the neutron-rich region, transitions in shape from spherical to prolate, prolate to oblate, 
and then oblate to spherical were observed. This indicates a transitional 
behavior of osmium isotopes. 
In Os nuclei ranging from N = 112 to N = 124, both the experimental data and our calculated results based on DD-ME1, DD-ME1, DD-PC1, DD-PCX, and NL3* exhibit consistent trends. However, deviations arise when comparing these trends with predictions derived from HFB theory utilizing SKM* functionals. 
It is important to note that nuclei near 
the drip-line region exhibit a spherical shape.

%%%%%%%%%%%%%%%%%%%%%%%%%%%%%%%%%%%%%%%%%%%%%%%%%5

\begin{landscape}
	
	\begin{longtable}{cccccccccc}
		\caption{Comparison of the Quadrupole Deformation Parameter ($\beta_2$) for the ground states of $^{158-266}$Os isotopes calculated using different effective interactions (DD-ME1, DD-ME2, DD-PC1, DD-PCX, and NL3*). The results are compared with available experimental data~\cite{22, wang2012, 43} and calculations performed using HFB models~\cite{19,39}}\label{tab:Osbeta}\\
		%		\hline
		\small
		\textrm{Nuclei}&
		\textrm{DD-ME1}&
		\textrm{DD-ME2}&
		\textrm{DD-PC1}&
		\textrm{DD-PCX}&
		\textrm{NL3*}&
		\textrm{EXP}&
		\textrm{SkP}&
		\textrm{SkM*}&
		\textrm{SLy4}\\
		%		\midrule   
		$^{156}$Os 	&	0.067	&	0.070	&	0.043	&	0.064	&	-0.077	&	--	&	--	&	--	&	--	\\
		$^{158 }$Os 	&	0.000	&	0.000	&	0.000	&	0.000	&	0.000	&	--	&	--	&	--	&	--	\\
		$^{160 }$Os 	&	-0.065	&	-0.067	&	-0.058	&	-0.060	&	0.000	&	--	&	--	&	--	&	--	\\
		$^{162 }$Os 	&	0.110	&	0.111	&	0.107	&	0.114	&	0.077	&	--	&	-0.085	&	0.074	&	0.112	\\
		$^{164 }$Os 	&	0.134	&	0.134	&	0.136	&	0.145	&	-0.092	&	--	&	0.141	&	0.124	&	0.145	\\
		$^{166 }$Os 	&	0.157	&	0.157	&	0.158	&	0.164	&	0.139	&	--	&	0.166	&	0.150	&	0.165	\\
		$^{168 }$Os 	&	0.167	&	0.168	&	0.171	&	0.177	&	0.164	&	--	&	0.187	&	0.169	&	0.179	\\
		$^{170 }$Os 	&	0.173	&	0.174	&	0.181	&	0.190	&	0.187	&	--	&	0.208	&	0.186	&	0.195	\\
		$^{172 }$Os 	&	0.185	&	0.187	&	0.196	&	0.208	&	0.305	&	0.224	&	0.266	&	0.201	&	0.278	\\
		$^{174 }$Os 	&	0.327	&	0.326	&	0.326	&	0.319	&	0.323	&	0.246	&	0.284	&	0.218	&	0.294	\\
		$^{176 }$Os 	&	0.338	&	0.336	&	0.337	&	0.331	&	0.331	&	--	&	0.288	&	0.260	&	0.302	\\
		$^{178 }$Os 	&	0.332	&	0.331	&	0.334	&	0.332	&	0.329	&	0.246	&	0.276	&	0.265	&	0.300	\\
		$^{180 }$Os 	&	0.322	&	0.320	&	0.322	&	0.307	&	0.319	&	0.248	&	0.256	&	0.242	&	0.278	\\
		$^{182 }$Os 	&	0.303	&	0.299	&	0.302	&	0.284	&	0.304	&	0.236	&	0.240	&	0.225	&	0.251	\\
		$^{184 }$Os 	&	0.286	&	0.284	&	0.285	&	0.263	&	0.287	&	0.208	&	0.226	&	-0.188	&	0.236	\\
		$^{186 }$Os 	&	0.264	&	0.262	&	0.261	&	0.240	&	0.269	&	0.200	&	0.214	&	-0.176	&	0.223	\\
		$^{188 }$Os 	&	0.243	&	0.241	&	0.234	&	0.226	&	0.245	&	0.184	&	0.200	&	-0.166	&	0.207	\\
		$^{190 }$Os 	&	0.190	&	0.190	&	0.191	&	0.193	&	0.202	&	0.177	&	0.184	&	-0.157	&	0.187	\\
		$^{192 }$Os 	&	0.167	&	0.167	&	0.168	&	0.167	&	0.174	&	0.164	&	0.166	&	-0.145	&	0.162	\\
		$^{194 }$Os 	&	0.149	&	0.149	&	0.150	&	0.148	&	0.150	&	--	&	-0.150	&	-0.129	&	-0.143	\\
		$^{196 }$Os 	&	0.121	&	0.122	&	0.119	&	0.119	&	0.124	&	--	&	-0.132	&	-0.111	&	-0.122	\\
		$^{198 }$Os 	&	-0.092	&	-0.092	&	-0.094	&	-0.094	&	-0.095	&	--	&	-0.104	&	-0.088	&	-0.096	\\
		$^{200 }$Os 	&	-0.072	&	-0.070	&	-0.072	&	-0.070	&	-0.064	&	--	&	-0.066	&	0.000	&	-0.061	\\
		$^{202 }$Os 	&	0.000	&	0.000	&	0.000	&	0.000	&	0.000	&	--	&	0.000	&	0.000	&	0.000	\\
		$^{204 }$Os 	&	0.000	&	0.000	&	0.000	&	0.000	&	0.002	&	--	&	0.000	&	0.000	&	0.000	\\
		$^{206 }$Os 	&	-0.055	&	-0.057	&	-0.050	&	-0.055	&	0.071	&	--	&	-0.033	&	0.000	&	0.024	\\
		$^{208 }$Os 	&	0.109	&	0.110	&	0.100	&	0.106	&	0.117	&	--	&	-0.046	&	0.000	&	0.076	\\
		$^{210 }$Os 	&	0.135	&	0.135	&	0.127	&	0.130	&	0.144	&	--	&	-0.064	&	0.000	&	0.120	\\
		$^{212 }$Os 	&	0.155	&	0.154	&	0.148	&	0.149	&	0.163	&	--	&	0.134	&	0.116	&	0.143	\\
		$^{214 }$Os 	&	0.170	&	0.169	&	0.182	&	0.170	&	0.178	&	--	&	0.165	&	0.142	&	0.166	\\
		$^{216 }$Os 	&	0.252	&	0.253	&	0.251	&	0.242	&	0.189	&	--	&	0.188	&	0.162	&	0.217	\\
		$^{218 }$Os 	&	0.279	&	0.278	&	0.275	&	0.268	&	0.299	&	--	&	0.234	&	0.180	&	0.243	\\
		$^{220 }$Os 	&	0.299	&	0.299	&	0.293	&	0.286	&	0.326	&	--	&	0.252	&	0.196	&	0.258	\\
		$^{222 }$Os 	&	0.305	&	0.303	&	0.300	&	0.290	&	0.329	&	--	&	0.262	&	0.217	&	0.266	\\
		$^{224 }$Os 	&	0.311	&	0.308	&	0.307	&	0.294	&	0.332	&	--	&	0.265	&	0.236	&	0.270	\\
		$^{226 }$Os 	&	0.318	&	0.315	&	0.313	&	0.300	&	0.338	&	--	&	0.263	&	0.240	&	0.270	\\
		$^{228 }$Os 	&	0.319	&	0.316	&	0.313	&	0.300	&	0.339	&	--	&	0.255	&	0.238	&	0.267	\\
		$^{230 }$Os 	&	0.318	&	0.314	&	0.312	&	0.299	&	0.340	&	--	&	0.245	&	0.233	&	0.263	\\
		$^{232 }$Os 	&	0.311	&	0.308	&	0.304	&	0.292	&	0.339	&	--	&	0.237	&	0.225	&	0.256	\\
		$^{234 }$Os 	&	0.296	&	0.293	&	0.289	&	0.275	&	0.329	&	--	&	0.230	&	0.216	&	0.245	\\
		$^{236 }$Os 	&	0.281	&	0.276	&	0.275	&	0.259	&	0.310	&	--	&	0.220	&	0.207	&	0.231	\\
		$^{238 }$Os 	&	0.266	&	0.263	&	0.263	&	0.248	&	0.292	&	--	&	0.208	&	0.197	&	0.217	\\
		$^{240 }$Os 	&	0.252	&	0.249	&	0.249	&	0.234	&	0.274	&	--	&	0.194	&	0.187	&	0.205	\\
		$^{242 }$Os 	&	0.239	&	0.236	&	0.235	&	0.219	&	0.260	&	--	&	0.182	&	0.177	&	0.192	\\
		$^{244 }$Os 	&	0.224	&	0.222	&	0.219	&	0.204	&	0.251	&	--	&	0.170	&	-0.151	&	0.177	\\
		$^{246 }$Os 	&	-0.150	&	-0.151	&	0.186	&	-0.157	&	0.229	&	--	&	-0.162	&	-0.141	&	0.162	\\
		$^{248 }$Os 	&	-0.137	&	-0.137	&	0.157	&	-0.142	&	-0.147	&	--	&	-0.150	&	-0.131	&	0.146	\\
		$^{250 }$Os 	&	-0.120	&	-0.121	&	0.138	&	-0.130	&	-0.128	&	--	&	-0.134	&	-0.120	&	0.125	\\
		$^{252 }$Os 	&	-0.100	&	-0.101	&	0.113	&	-0.117	&	-0.104	&	--	&	-0.119	&	-0.106	&	-0.115	\\
		$^{254 }$Os 	&	0.083	&	0.086	&	-0.097	&	0.084	&	-0.094	&	--	&	-0.102	&	-0.088	&	-0.100	\\
		$^{256 }$Os 	&	0.058	&	0.062	&	-0.066	&	0.018	&	0.085	&	--	&	-0.079	&	0.000	&	-0.074	\\
		$^{258 }$Os 	&	0.000	&	0.000	&	0.000	&	0.000	&	0.069	&	--	&	-0.039	&	0.000	&	0.000	\\
		$^{260 }$Os 	&	0.000	&	0.000	&	0.000	&	0.000	&	0.031	&	--	&	0.000	&	--	&	--	\\
		$^{262 }$Os 	&	0.002	&	0.002	&	0.001	&	0.004	&	0.058	&	--	&	0.450	&	--	&	--	\\
		$^{264 }$Os 	&	0.005	&	0.005	&	0.003	&	0.006	&	0.093	&	--	&	0.447	&	--	&	--	\\
		$^{266 }$Os 	&	0.008	&	0.008	&	0.006	&	0.007	&	0.125	&	--	&	--	&	--	&	--	\\
		\hline
	\end{longtable}		
\end{landscape}

%%%%%%%%%%%%%%%%%%%%%%%%%%%%%%%%%%%%%%5
%%%%%%%%%%%%%%%%%%%%%%%%%%%%%%%%%%%%%%%%
%%%%%%%%%%%%%%%%%%%%%%%%%%%%%%%%%%%%%%%%%%%%55
\begin{figure}[hbt!]
	\begin{center}
		\includegraphics[scale=0.45]{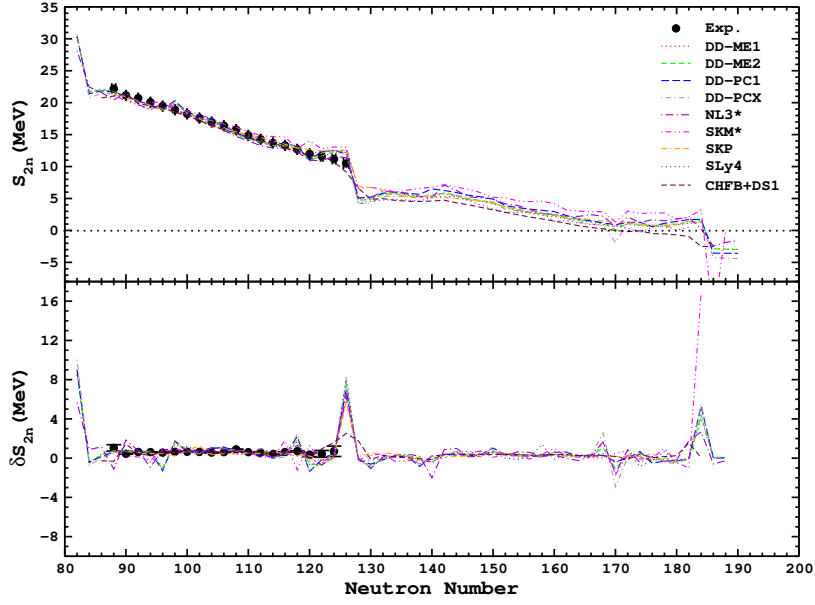}
		\caption{Two-neutron separation energies (upper panel), and two-neutron shell 
			gaps (lower panel), for even-even Osmium nuclei from N=82 to neutron N = 190, obtained using different interactions. 
			The other interactions and results are presented for comparison.}
		\label{Ossn}
	\end{center}
\end{figure}

\begin{figure}[hbt!]
	\begin{center}
		\includegraphics[scale=0.45]{epair_Os.eps}
		\caption{The neutron pairing energy (E$_{pair,n}$ ) for even-even Osmium nuclei from N=82 
			to N = 190, obtained using different interactions.}
		\label{Osep}
	\end{center}
\end{figure}
%%%%%%%%%%%%%%%%%%%%%%%%%%%%%%%%%%%%%%%5

\subsection{Two-neutron separation energy ($S_{2n}$), 
	shell gap $\delta S_{2n}$ and neutron pairing energy (E$_{pair,n}$)}
The identification of shell closures in atomic nuclei is crucial for understanding nuclear structure. Two important observables used in this regard are the two-neutron separation energy($S_{2n}$) and the two-neutron shell gap ($\delta S_{2n}$). The $\delta S_{2n}$, also known as the $S_{2n}$ differential, is calculated using the following relation:
$$\delta S_{2n} = \frac{S_{2n}(N,Z) - S_{2n}(N+2,Z)}{2}$$
We present S$_{2n}$, and $\delta S_{2n}$ for even-even $^{158-266}$Os isotopes in Fig.~\ref{Ossn}.  
To validate our calculations, we compare our results with available experimental data and also with predictions from other theoretical models. Specifically, we compare our Relativistic Mean Field (RMF) values with those obtained from the HFB+THO model using Skyrme SLy4, SkP, and SkM* functionals~\cite{19}, as well as the CHFB+5DCH model based on the Gogny D1S interaction~\cite{39}.

In Fig.~\ref{Ossn}, we observe a sudden decrease in the two-neutron separation energy (S$_{2n}$) and pronounced kinks in the $\delta S_{2n}$ at neutron numbers N=82, N=126, and N=184. These findings are consistent with the long-established neutron shell closure at N=82, N=126, and N=184. Additionally, we note the presence of kinks at N=118 and N=168 in the relativistic interactions (DD-ME1, DD-ME2, DD-PC1, DD-PCX, NL3*), suggesting a possible neutron shell or subshell closure at these neutron numbers. However, non-relativistic functionals such as SkP, SkM*, SLy4, and CHFB do not exhibit such kinks at N=118 and and N=168. Previous research has proposed the existence of several deformed subshell closures~\cite{Bassem, Naz2018,Naz2019,afaque}.

The idea of neutron shell closure based on the disappearing neutron pairing energy~\cite{Del2001,Sil2004}. Supporting the notion of shell closures, the neutron pairing energy (E$_{pair,n}$) vanishes at neutron numbers 82, 126, and 184, as shown in Fig.~\ref{Osep}. We found that E$_{pair,n}$ becomes zero at N = 118, matching the expected shell behavior seen in Fig.~\ref{Ossn} with $S_{2n}$ and $\delta S_{2n}$. 
However, at N = 168, E$_{pair,n}$ does not reach zero, which challenges the idea of a shell closure at this neutron number. This discrepancy suggests that the shell structure at N = 168 is different from what is typically seen at closed neutron shells. This highlights the complexity of nuclear shell evolution, especially in heavy, neutron-rich nuclei, and suggests the need for further study.
These observations provide valuable insights into the nuclear structure of Os isotopes and contribute to our understanding of shell closures in atomic nuclei.

%%%%%%%%%%%%%%%%%%%%%%%

\begin{figure}[hbt!]
	\begin{center}
				\includegraphics[scale=0.45]{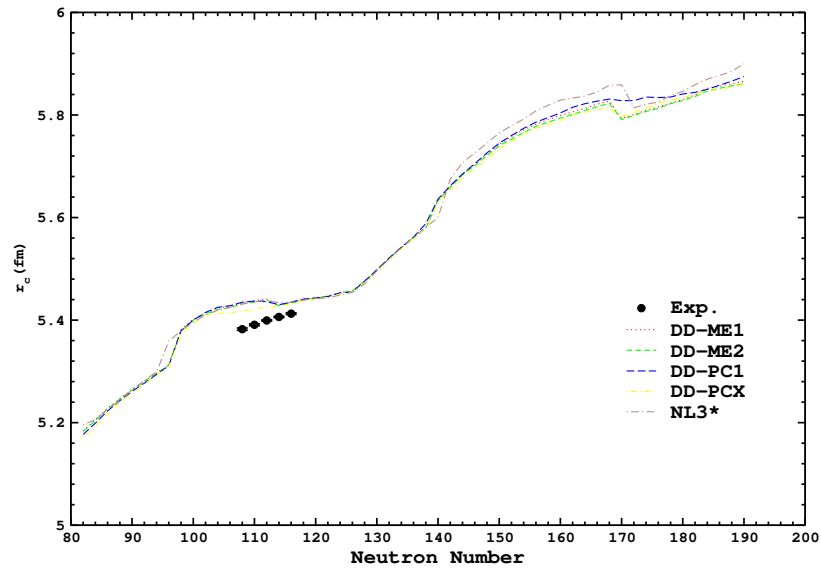}
		\caption{The charge radii (r$_c$) for even-even Osmium nuclei from N=82 
			to N=190, obtained using different interactions.}
		\label{Osrc}
	\end{center}
\end{figure}
%%%%%%%%%%%%%%%%%%%%%%%
\begin{figure}[hbt!]
	\begin{center}
				\includegraphics[scale=0.45]{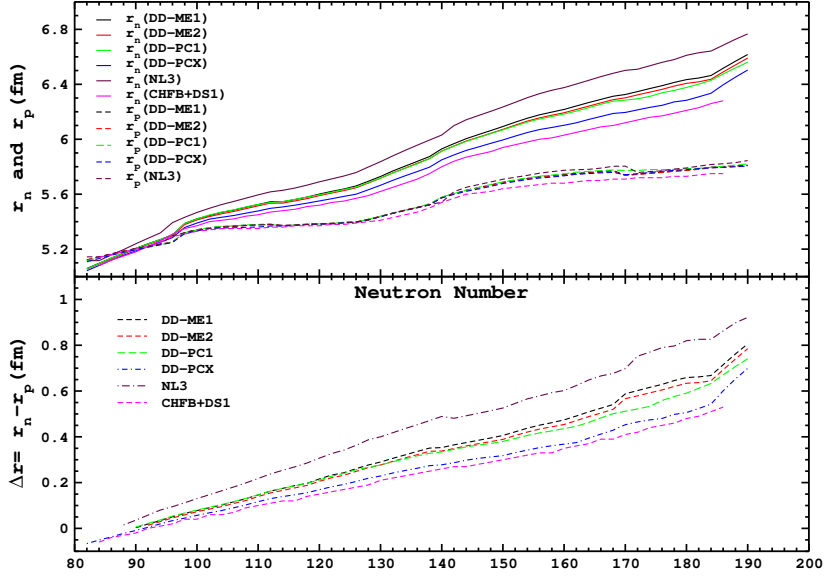}
		\caption{The nucleon radii($r_{n}$, and $r_{p}$) (upper-panel)
			and neutron skin thickness ($\Delta r=r_{n}-r_{p} $) 
			(lower-panel) for even-even Osmium nuclei from N=82 
			to the neutron drip line, obtained using different interactions.}
		\label{Osrad}
	\end{center}
\end{figure}
%%%%%%%%%%%%%%%%%%%%%%%

\subsection{Neutron, proton and charge radii}
Nuclear radii provide crucial information about the size and distribution of neutrons and protons within a nucleus. In particular, the neutron skin thickness, which is the difference between the neutron and proton radii, becomes significant in neutron-rich nuclei where an excess of neutrons is present. The neutron skin thickness is directly related to fundamental properties of nuclear matter and helps bridge the understanding of finite nuclear systems to infinite nuclear matter~\cite{Reinhard2016,Hagen2015,Agarwal2006}.

In this subsection, we aim to investigate the correlation between nuclear shape and the distribution of nuclear matter. We compare theoretical results obtained from different models and also compare them with available experimental data. The charge radii ($r_c$) are presented in Fig.~\ref{Osrc}, while the root-mean-square (RMS) nucleon radii ($r_n$ and $r_p$) and the neutron skin thickness ($\Delta r = r_n - r_p$) for the considered isotopic chains are shown in Fig.~\ref{Osrad}.
For even-even Os isotopes, the neutron radius exhibits an increasing trend with an increasing number of neutrons. The dips observed at neutron numbers N=82, 126, and 184 indicate the closed shell behavior of these nuclei. Due to variations in deformation, the increments in different radii for even-even Os isotopes are not entirely smooth.
We compare our calculated values of the RMS charge radii ($r_c$) with the available experimental data~\cite{angeli2004,31}. The comparison reveals a good agreement between our theoretical results and experimental measurements.
Fig.~\ref{Osrad} demonstrates that the neutron skin thickness ($\Delta r$) monotonically increases with the number of neutrons. This observation further supports our understanding of the relationship between neutron-rich nuclei and their matter distribution.

\subsection{Potential energy curve}
\begin{figure*}%[hbt!]
	\centering
				\includegraphics[scale=0.5]{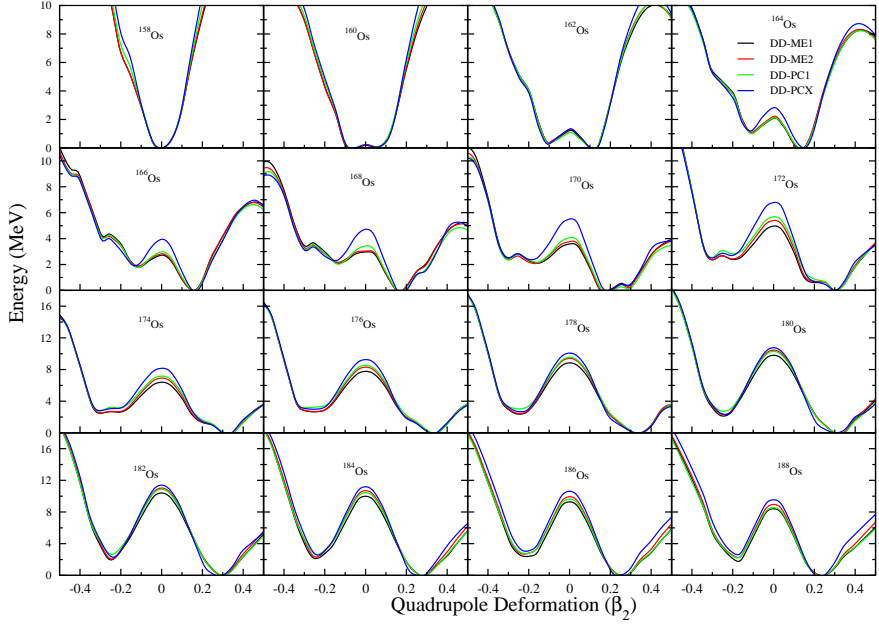}
	\caption{Potential energy curves for the even-even isotopes 
		$^{158-188}$Os as functions of the quadrupole deformation from 
		DD-ME1, DD-ME2, DD-PC1, and DD-PCX force parameters. 
		The curves are scaled with the ground state binding energy.}
	\label{ps1}
\end{figure*}

\begin{figure*}[t]
	\begin{center}
				\includegraphics[scale=0.5]{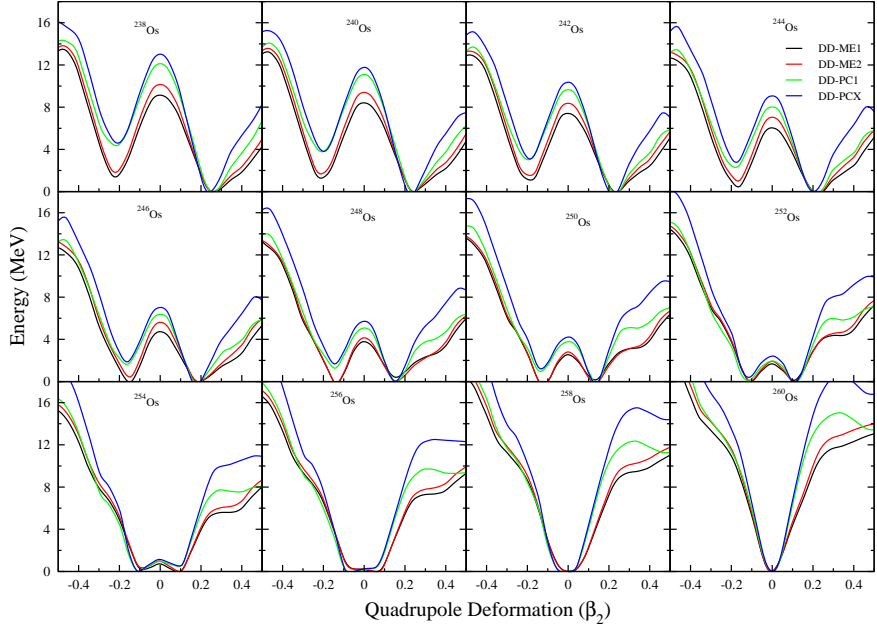}
		\caption{Same as Fig.~\ref{ps1} but for $^{238-260}$Os.}
		\label{ps4}
	\end{center}
\end{figure*}

The shape of a nucleus is one of its most essential as well as fundamental features, described 
by quadrupole deformation.
Most of the nuclei exhibit the spherical as well as ellipsoid shape. 
In axially symmetric case, a deformed nucleus is classified 
by quadrupole deformation parameter $\beta_2$. 
Here, we employed CDFT with density-dependent effective interactions 
DD-ME1~\cite{me1}, DD-ME2~\cite{me2}, DD-PC1~\cite{pc1}, and 
DD-PCX~\cite{pcx} to perform quadrupole-constrained calculations for 
examining the Potential energy curves for even-even Os chain. 
The potential energy curves for the considered isotopic chain are plotted in Figs.~\ref{ps1}-~\ref{ps4}.

\begin{figure*}%[hbt!]
	\begin{center}
				\includegraphics[scale=0.5]{pse_Os2.eps}
		\caption{Same as Fig.~\ref{ps1} but for $^{190-212}$Os}
		\label{ps2}
	\end{center}
\end{figure*}

\begin{figure*}
	\begin{center}
				\includegraphics[scale=0.5]{pse_Os3.eps}
		\caption{Same as Fig.~\ref{ps1} but for $^{214-236}$Os.}
		\label{ps3}
	\end{center}
\end{figure*}

One can see that the results do not differ from density-dependent 
meson exchange model to point coupling model. 
It can also be noted that the location of spherical, oblate, and prolate
minima appears at the same deformation, no matter what the force or model is used. 
However, the relative energies can be slightly different depending on the models.  
In general, the outcomes of energy barriers for deformations through 
all the effective interactions are found to be comparable. 
The ground state of $^{158}$Os is revealed to be spherical confirming the 
spherical form as expected at well-known magic neutron number N = 82. 
As the neutron number increases, the global minima move 
from spherical to prolate region. 
$^{162}$Os and $^{164}$Os seem to exist in both slightly prolate 
and oblate shapes as the corresponding minima have an energy difference 
of about 1.0 MeV or less. 
In $^{166-194}$Os isotopes, we see oblate shapes with an excited 
state at about $2.5$ MeV higher than the global minima. 
Additionally, Figs.~\ref{ps2} clearly shows that there is a very strong 
competition between different low-lying configurations corresponding 
to different intrinsic deformations, and this situation of a nucleus 
is described by shape coexistence.
Here, in considered nuclei, $^{196}$Os nucleus has two separate minima 
at $\beta_2=0.1$ (prolate) 
and $\beta_2=-0.1$ (oblate) with a small amount of energy difference about 0.5 MeV.

It is worth mentioning that a clear shape evolution from 
spherical to prolate as well as oblate shapes is observed for increasing neutron number. 
A shape transition can be seen from deformed to spherical through flat minima. 
The ground state of $^{202}$Os has a sharp global minimum at $\beta_2=0.0$
refers to N = 126 neutron magicity.
Again in the neutron-rich region, the sphericity disappears and the 
global minima move to the well-deformed prolate side. 
The $^{214-238}$Os isotopes show well-deformed prolate shapes, and the 
prolate minima lie $2.0-3.0$ MeV deeper than the oblate minima in the case of 
DD-ME1 and DD-ME2 calculations while DD-PC1, DD-PCX interactions give 
$2.0-4.0$ MeV energy difference for prolate to oblate as a 
first intrinsic excited state.
The isotopes, $^{240-254}$Os coexist as both prolate and oblate in their ground 
state, while DD-PC1 and DD-PCX interactions are not affirmative 
regarding the coexistence but favor prolate as a ground state. 
One can see that spherical minimum for $^{260}$Os supports the N = 184 
as a neutron magicity.
Our results are in qualitatively good agreement
with those earlier predictions obtained in Refs.~\cite{afana2016,Sarriguren2008,Fossion2006}.

%%%%%%%%%%%%%%%%%%%%%%%%%%%%%%%%

\section{Conclusion}
In this study, we provide a comprehensive analysis of the Os isotopic chain, ranging from the beta-stable region to the neutron drip-line region, using a covariant density functional and relativistic self-consistent mean-field description. Various ground state bulk properties, including binding energy, two-neutron separation energy ($ S_{2n}$), two-neutron shell gap 
($\delta S_{2n}$), neutron pairing energy (E$_{pair,n}$), quadrupole deformation parameter, 
rms radii, and neutron skin thickness ($\Delta r$) are calculated.
The results obtained from the covariant density functional theory (CDFT) and relativistic mean-field (RMF) approach are compared with the self-consistent Hartree-Fock-Bogoliubov (HFB) formalism based on the Skyrme SLy4, SkP, and SkM* interactions~\cite{19}, as well as the extended HFB theory (CHFB+5DCH) with the Gogny D1S interaction~\cite{39}. These calculations are further compared with available experimental data~\cite{22,33,34,31,angeli2004}, and a good agreement is observed between them.

In the isotopic series under investigation, prominent shell closures are observed at N = 82, 126, and 184, while N = 118 is suggested to be a shell or subshell closure. These closures are identified based on the behavior of two-neutron separation energy, two-neutron shell gap, and neutron pairing energy. The quadrupole deformation parameters reveal shape transitions across the isotopic series. Additionally, significant evidence of shape coexistence is found in in $^{196}$Os and $^{248-254}$Os, which is consistent with the HFB results.

In order to investigate the phase shape transition in the even-even osmium isotopic series, we employ the Covariant Density Functional Theory (CDFT) with separable pairing. Our calculations utilize different parameter sets, namely the density-dependent DD-ME1~\cite{me1}, DD-ME2~\cite{me2}, 
DD-PC1~\cite{pc1}, DD-PCX~\cite{pcx}, and nonlinear NL3*~\cite{nl3s} interactions. The ground state configurations are determined by locating the minima on the potential energy curve, which allows us to study phenomena such as shape coexistence and phase shape transitions.
 
It is worth noting that the nuclear shape transitions primarily occur at shell closures, indicating the transitional nature of osmium isotopes. Notably, the meson exchange and point coupling interactions (i.e., DD-ME1, DD-ME2, DD-PC1, DD-PCX) yield consistent behavior in the evolution of shape within the potential energy curves. The analysis of these results leads us to the conclusion that the potential energy curves are insensitive to the choice of interactions used in the Density Functional Theory.
 
The results obtained from CDFT using various effective interactions provide a reasonable description of the ground-state bulk properties. Furthermore, these findings exhibit strong similarities with non-relativistic Hartree-Fock-Bogoliubov (HFB) calculations and other theoretical predictions. Consequently, we can confidently state that the outcomes presented in this study are independent of the specific models and force parameters employed, thus highlighting the authenticity and robustness of our predictions.

\section*{Acknowledgements}
One of the authors (UR) would like to thank the Department of Physics, Aligarh Muslim 
University, for using the computational facility.

\bibliographystyle{spphys}       % APS-like style for physics
\bibliography{sn-bibliography}   % name your BibTeX data base

\end{document}